\documentclass{elsart}
\usepackage{natbib}
\usepackage{graphicx}

\begin{document}

\runauthor{E. Del Gado and W. Kob}

\begin{frontmatter}

\title{
Network formation and relaxation dynamics in
a new model for colloidal gelation. 
}
\author[1]{Emanuela Del Gado}
\author[2]{Walter Kob}
\address[1]{ETH Z\"urich, Department of Materials, Polymer Physics, 
HCI H 541, CH-8092 Z\"urich, Switzerland}
\address[2]{Laboratoire des Collo\"\i des, Verres et Nanomat\'eriaux, 
UMR5587 CNRS, Universit\'e Montpellier 2, 34095 Montpellier, France}

\begin{abstract}
We investigate the gel formation from the equilibrium sol phase in a 
simple model that has the characteristics of (colloidal) gel-forming 
systems at a finite temperature. 
At low volume fraction and low temperatures, particles are linked 
by long-living bonds and form an open percolating network. 
By means of molecular dynamics simulations, we study the lifetime of 
bonds and nodes of the gel network in order to relate these quantities 
to the complex relaxation dynamics observed. 
\end{abstract}
\end{frontmatter}

\section{Introduction}
Gels are a special case of slow and disordered systems: Due to the low volume
fraction, their slow dynamics are intimately related to the formation of an
open structure~\cite{exp}. 
The understanding of the connection between the structure of
these systems and their complex dynamics is fundamental in the research on
complex fluids, such as biological systems, polymeric materials, and colloids.
The slow dynamics of gelling systems have some analogies with the dynamics of
other disordered systems such as dense glasses, in that the 
majority of the particles becomes temporarily localized in space due 
to the structure formation.
On the other hand, due to the presence of the network, the gels show  
relaxation dynamics which strongly depend on the length scale/wave 
vector considered. 
At small wave vectors and close to the gelation threshold one 
typically observes very slow stretched-exponential and power law relaxations. 
Surprisingly, at higher wave vectors, recent experimental work on colloidal 
gels shows the onset of a compressed exponential relaxation 
(i.e. Kohlraush exponent $\beta >1$)
~\cite{compr}.
This non-trivial behavior is the subject of an intense debate in the last
years. In particular, it is still not clear how is it related to the structural
features of the gel.\\
Numerical simulation studies offer an effective tool in that they allow to
follow the local relaxation processes and to understand their role in the
complex dynamics. In spite of this, progress has also been hampered by the
fact that there are so far very few microscopic realistic models for gels that
effectively allow to study their dynamics by means of analytical methods or 
computer simulations. In colloidal systems the interparticle interactions 
can be suitably tuned: First, particles can be coated and stabilized 
leading to a hard sphere behavior. Then, an attractive depletion interaction 
can be induced by adding some non-adsorbing polymers. 
The range and strength of the potential are controlled by the size and 
concentration of the polymer respectively~\cite{edinburgh}.
For volume fractions from 3\% up to 40\% and attractive interaction energies 
of the order of $k_BT$ confocal 
microscope images~\cite{conf} 
show the formation of persistent structures, different at 
different volume fractions, and gel phases, whose mechanisms of formation are 
still poorly understood. It is important to note that here 
the underlying thermodynamics may significantly interplay and/or compete
with gel formation via phase separation or microphase separation. 
As a consequence, coarsening or ordering processes will interfere with 
the gel dynamics. Whereas in the experiments the time scales typical 
of the micro or macro-phase separation are often much longer than the 
observation time scales, this is not the case in numerical studies using
traditional models for colloidal suspensions, where the investigation of the
gel dynamics has been severely hindered until very recently.\\
In order to overcome these difficulties, we have recently proposed a new 
model~\cite{delgado_05a,delgado_05b} consisting of meso-particles 
which interact via a short range attraction of the type already used 
to study colloidal systems. We have started from the idea that, in order to 
obtain a persistent open structure, a mechanism competing with phase separation
processes is needed.
The novelty of our model is to include a directional effect in the
interactions and therefore to facilitate the formation of the open structures
found in gelling systems. These directional interactions are given by a
combination of site-site interactions of the particles as well as a short
ranged three-body interaction. Thus this new approach will promote the formation
of large-scale disordered structures that are stable also at relatively high
temperatures, without imposing a fixed local functionality of the meso-particle.
It should also be noted that the presence of directional effective 
interactions, although not yet thoroughly investigated, is likely to play an 
important role in the physics of colloidal gels. This is supported 
by the experimental observation of the local rigidity of the structures 
and of their low local connectivity: The open structures which spontaneously 
form at low volume fractions are characterized by aggregating units 
(particles or aggregates of particles) with a coordination number between 3 
and 4~\cite{dinswei}. In addition, surface inhomogeneities or sintering 
of the particles which are present in many cases can produce quite rigid 
links~\cite{books}.\\ 
In this paper, 
we analyze the formation of the gel network in this model and focus on the
time autocorrelation of bonds and network nodes. This study is essential to
make the connection between the features of the network structure and the 
onset of the complex dynamics of the gel phase \cite{delgado_05b}.\\
In the following we briefly present the model and the numerical study
in Sec.~\ref{mod}. We then describe the aggregation process and the structure
of the gel phase in Sec.~\ref{str}. In Sec.~\ref{lifetime} 
we study the time autocorrelation functions and the lifetime of bonds and of 
the nodes of the network structure. Finally, conclusions drawn from the 
results of the numerical study are contained in Sec.~\ref{conclu}.

\typeout{SET RUN AUTHOR to \@runauthor}
\section{Model and numerical simulations}
\label{mod}
The model investigated consists of identical particles of radius 
$\sigma$, interacting via an effective potential $V_{\rm eff}$ that is 
the sum of a two-and three body terms. The two body potential is itself 
the sum of a hard core like interaction that is given by a generalized 
Lennard-Jones potential, $V_{\rm LJ}(r)$, and a term $V_{\rm cp}$ that
depends on the relative orientation of the particles. For the radial
term we have used 
$V_{\rm LJ}(r)= 23 \epsilon [(\sigma/r)^{18}-(\sigma/r)^{16}]$
where the prefactors and exponents have been chosen in such a way to
give a relatively narrow well of depth $\epsilon$.
In the following we will measure length and energy such that $\sigma
=0.922$ and $\epsilon=1$, respectively, and time in units of $\sqrt{m
\sigma^2 /\epsilon}$, where $m$ is the mass of a particle.\\ 
As already discussed in the previous section, we have introduced a mechanism
that favors the formation of an open network structure, in competition with
the phase separation induced by purely radial interactions. This has allowed us 
to obtain an open persistent structure at temperatures and volume fractions 
where phase separation or microphase separation do not occur and therefore to 
study the dynamics of the gel formation without the complex interplay 
with the underlying thermondynamics. In practice, we have introduced a 
directional interaction by decorating each particle with 12 points that 
form a rigid icosahedron inscribed in a sphere of diameter $1.1\sigma$. 
The potential $V_{\rm cp}$ between a particles \#1 with a particle 
\#2 is then set up in such a way that it is more favorable that the center 
of particle \#1 approaches particle \#2 in the direction of one of the points 
of the icosahedron that decorate particle \#2. In addition we have also 
included an explicit (short range) three body potential $V_3$ in the form 
of a gaussian in the angle $\theta$ between three neighboring particles and 
which makes that values of $\theta$ smaller than $0.4 \pi$ are unlikely. 
Although recently 3-body interactions have been directly measured
amongst charged colloidal particles~\cite{3b}, one could more generally see 
the particles in our model as the building blocks of the gel network 
(i.e. particles or aggregates of particles). In this terms, our model is aimed 
to describe the presence of directional effective interactions at some 
mesoscopic length scale, leading to the formation of the open structures.
More details on these potentials will be given elsewhere~\cite{delgado_05c}, 
but here it should be noted that, whereas $V_{3}$ is a real 
3-body term introducing an angular rigidity, 
$V_{\rm cp}$ is a geometric term which 
allows to add a soft-sphere repulsion depending on relative orientation. 
This term is not able alone to effectively limit the functionality of the 
particles at the volume fraction and temperatures considered here.
The choice of considering the effect of both these terms has been
made in the spirit of investigating more deeply their relative 
contribution to the formation of the open network (and to its dynamics).
For example, by taking only $V_{LJ} + V_3$ an open structure can certainly 
be obtained, but probably slightly different from the one discussed here. 
For the moment a complete understanding of the interplay of these two terms 
has not been reached yet.\\
On the whole, in our model the (meso) particles can form directional bonds
that favor the formation of an open network structure, without, however,
imposing a local symmetry or connectivity, in contrast to models that
have been proposed before~\cite{delgado_latt,zaccarelli2}.
As a consequence, in our case there is an {\em effective} 
functionality of the particles, which is the result of the competition 
between entropic and internal energy contribution in the network
formation.\\
Using these interactions, we have done microcanonical simulations
using constrained molecular dynamics and a suitable combination of 
the algorithms RATTLE and SHAKE~\cite{md} with a step size of 0.002.
The number of particles was 8000 and the size of the simulation
box $L=43.09$ which gives a volume fraction of 0.05. (This corresponds
to a particle density of 0.1)  Before starting the production runs
we carefully equilibrated the system by monitoring that the relevant
time correlation functions have attained their asymptotic limit. The
temperatures investigated were 5.0, 2.0, 1.0, 0.7, 0.5, 0.3, 0.2, 0.15,
0.1, 0.09, 0.08, 0.07, 0.06, and 0.05. In order to improve the statistics of
the results we have averaged them over five independent runs.

\section{Structure and cluster formation}
\label{str}
\begin{figure*}
\vspace*{1cm}
\begin{center}
\includegraphics[width=0.9\linewidth]{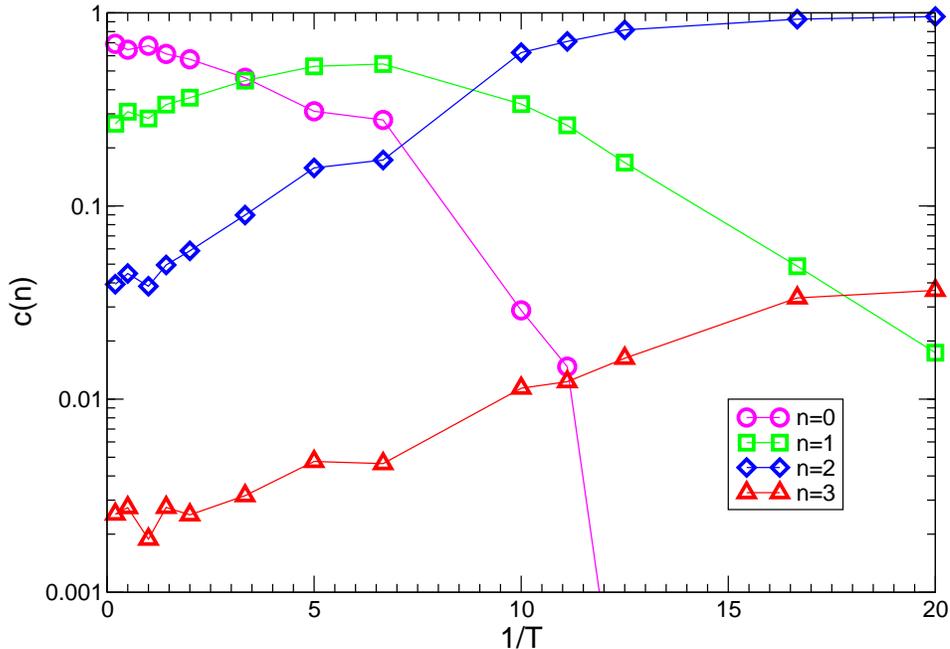}
\caption{(Color online)
 $T-$dependence of $c(n)$, the fraction of particles
having coordination number $n$.
}
\label{fig1}
\end{center}
\end{figure*}

The change of the topology of the structure with decreasing $T$ can be
characterized by investigating the coordination number $c(n)$, which is
shown in Fig.~\ref{fig1} as a function of the inverse temperature. 
We define $c(n)$ as the fraction of particles that have exactly $n$ 
neighbors. (Two particles are considered to be neighbors if their distance 
is less that $r_{\rm min}$=1.1, the location of the first minimum in the 
radial distribution function.) 
We see that at high temperatures a large majority of the particles are
isolated ($n=0$). By lowering the temperature, the number of these free 
particles rapidly decreases and they practically disappear. The fraction 
of dimers or chain ends, $n=1$,
is around 30\% at high $T$. With decreasing $T$ such fraction increases,
 it attains a maximum at
around $T=0.15$, and then decreases quickly with decreasing
$T$. 
At the same time the number of particles that have exactly two nearest neighbors
increases rapidly and these (local) configurations become by far the most 
prevalent ones at low $T$. 
Last not least also the number of particles with $n=3$
neighbors increases quickly with decreasing $T$. From these curves we thus
can conclude that with decreasing $T$ the system forms an open network in
which most particles form chains that meet at points with coordination
number three and which are thus important for the mechanical properties
of the structure. 
\begin{figure*}
\vspace*{1cm}
\begin{center}
\includegraphics[width=0.65\linewidth]{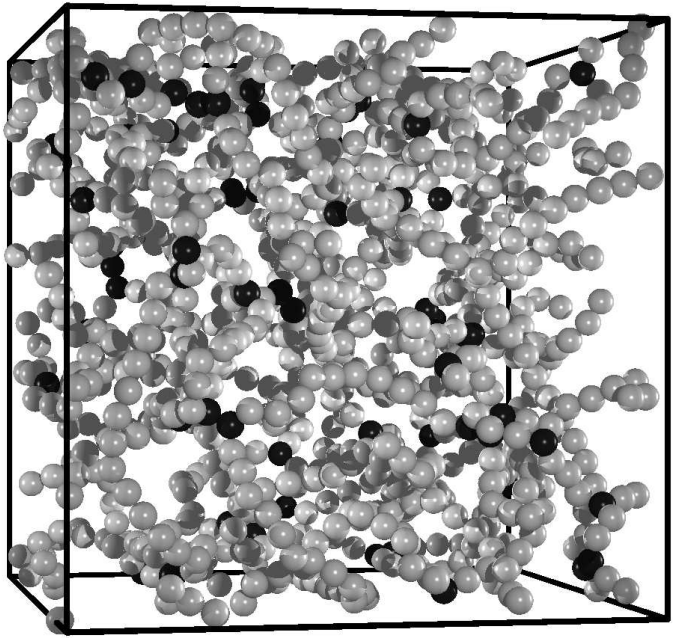}
\caption{
Part of a configuration, a cube with side $L/2$, of
the system at $T=0.05$.
The light and dark particles have coordination
$n=2$ and $n=3$, respectively.}
\label{fig2}
\end{center}
\end{figure*}
We have also shown elsewhere~\cite{delgado_05a,delgado_05b} 
that whereas at high $T$ the static structure factor $S(q)$ of the system 
is relatively flat, i.e. the system has a structure similar to the one 
of a gas of free particles, around $T=0.1$ $S(q)$ starts to show a 
peak at $q_0=7.7$, a wave-vector which corresponds
to the distance between two nearest neighbor particles.
In addition $S(q)$ shows an increase at small $q$ which indicates the
formation of a disordered open network structure. This increase
is relatively moderate and does not strongly depend on $T$, giving evidence
that the system does not undergo a phase separation even at the
lowest temperatures considered. Moreover, $S(q)$ does not show
any pronounced peak at these low wave-vectors and thus we can conclude
that the network is disordered and does not have a well defined length scale.\\
The appearance of the peak at $q_0=7.7$ is due to the fact that at
intermediate and low $T$ the particles condense into clusters.
By defining that  a particle belongs to a
cluster if its distance from at least one member of the cluster is less
than $r_{\rm min}$, we have studied $n(s)$, the number of clusters that
have exactly $s$ particles~\cite{delgado_05a}. 
For $T\geq 0.3$ this distribution follows an exponential law, corresponding to
the random formation of transient clusters of non-bonded particles at low
densities.
Coherently with the data shown in Fig.\ref{fig1},
at $T=0.1$ the shape of the distribution has strongly changed in that now
the most probable clusters have size $s=2$.
At the same time $n(s)$ has grown a tail at large $s$ in that the largest
clusters found have now $O(100)$ particles. At lower temperatures,
$n(s)$ crosses over to a power law regime for high values of $s$,
with a crossover point that moves to larger $s$ with decreasing $T$.
At T=0.06 this regime is apparently compatible with
random percolation~\cite{stauffer}. 
At the lowest temperatures, $T\leq 0.06$, the
distribution shows a gap at large $s$ in that the system can form one
big cluster that contains a substantial fraction of
the particles in the system. Finally at $T=0.05$ we have only very few
particles that are members of small clusters, with an $n(s)$ that is
basically a constant, whereas the overwhelming majority ($ > 97 \% $) 
of the particles belongs to one large percolating cluster. 
Hence, at very low temperatures it is energetically
very unfavorable to have free particles and small free clusters.\\
In Fig. \ref{fig2} a typical configuration at $T=0.05$ is shown:
An open network in which most particles are forming
chains (light spheres) that are connected in a relatively random way
at points that have three or more neighbors (dark spheres).

\section{Dynamics: Lifetimes of bonds and nodes of the network}
\label{lifetime}
\begin{figure*}
\vspace*{1cm}
\begin{center}
\includegraphics[width=0.9\linewidth]{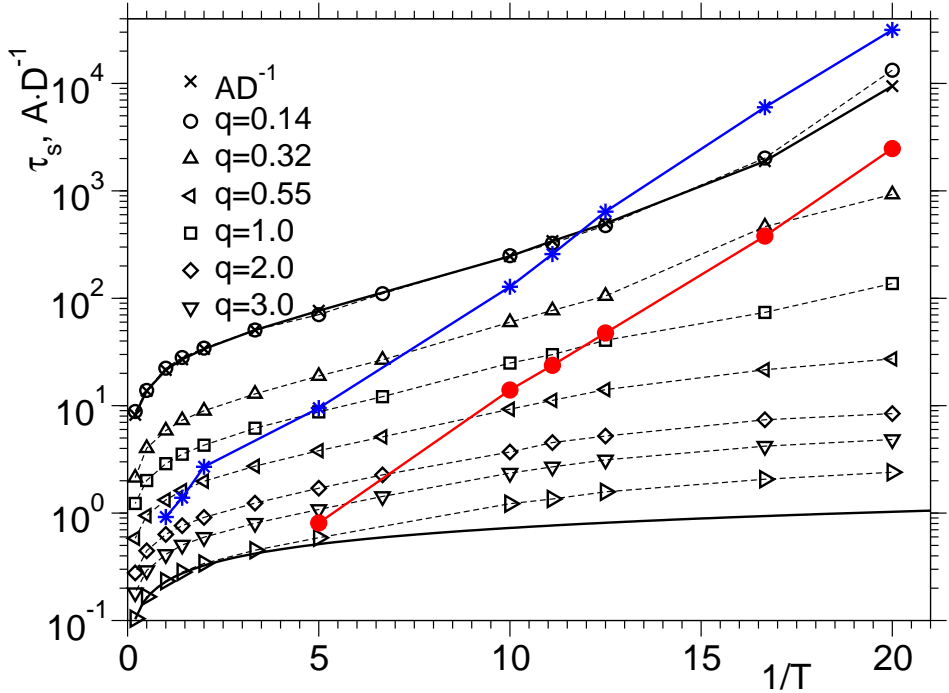}
\caption{(Color online) Arrhenius plot of the diffusion constant
$D$ and of the relaxation time $\tau_s(q,T)$ as determined from the
self-intermediate scattering function $F_s(q,t)$. The solid line is a
fit to the high $T$ data for $q=3.0$ of the form $\tau={\rm const.}/
\sqrt{T}$. The open symbols correspond to different wave-vectors. The stars
and filled circles are $\tau_b$ and $\tau_{3b}$, respectively.
}
\label{fig3}
\end{center}
\end{figure*}

Having characterized the structure of the system, we now turn our
attention to its dynamical properties.
We have studied the mean-squared displacement of a tagged particle, 
$\langle r^2(t) \rangle=\langle |{\bf r}_j(t)- {\bf r}_j(0) |^2 \rangle$.
A detailed discussion of the time dependence of $\langle r^2(t) \rangle$ 
can be found in Ref.\cite{delgado_05b}.
Using this function and the Einstein relation we obtained the diffusion 
constant $D(T)$. 
Furthermore we have calculated the self-intermediate scattering function 
$F_s(q,t)$ for wave-vector $q$, $F_s(q,t) = N^{-1} \sum_{j=1}^N 
\langle \exp[i {\bf q} \cdot ({\bf r}_j(t)- {\bf r}_j(0))]\rangle$.
The discussion of the quite complex $q$ and $T$ dependence of $\langle
r^2(t) \rangle$ and $F_s(q,t)$, characteristic for a gel-forming
system, has also been presented elsewhere~\cite{delgado_05b}. 
From these functions we have calculated the characteristic relaxation times
$\tau_s(q,T) = \int F_s(q,t) dt$.\\ 
Here we focus on the time autocorrelation functions of bonds and nodes 
forming the gel network. This analysis is essential to understand the
role of different parts of the structure in the complex dynamics observed.
As in Fig.\ref{fig2} we make the distinction of bonds 
that are connected to particles that have a coordination number of three
(referred to as ``anchor particles'') from the bonds that are connected 
to particles that have a coordination number of two (``bridging particles'').
For this we have determined $C_b(t)$, the probability that a bond that exists 
at time zero is still present at time $t$, defined as 
\begin{equation}
C_b(t) = \frac{\sum_{ij}
\left[ \langle n_{ij}(t)n_{ij}(0)\rangle-\langle n_{ij}\rangle^2\right]}%
{\sum_{ij}\left[\langle n_{ij}^2\rangle-\langle n_{ij}\rangle^2\right]},
\end{equation}
where $n_{ij}(t)=1$ if particles $i$ and $j$ are linked at time $t$ and
$n_{ij}(t)=0$ otherwise. 
We have also calculated the corresponding time correlation function for the 
anchor points $C_{3b}(t)$ defined as
\begin{equation}
C_{3b}(t) = \frac{\sum_{i}
\left[ \langle n_{3i}(t)n_{3i}(0)\rangle-\langle n_{3i}\rangle^2\right]}%
{\sum_{i}\left[\langle n_{3i}^2\rangle-\langle n_{3i}\rangle^2\right]},
\end{equation}
where $n_{3i}(t)=1$ if particles $i$ is an anchor point at 
time $t$, $n_{3i}(t)=0$ otherwise. 
We have determined the corresponding average lifetime
of bonds and anchor points as $\tau_b(T)= \int C_b(t) dt$ and
$\tau_{3b}(t) = \int C_{3b}(t) dt$, respectively.
In Fig.~\ref{fig3} these quantities are plotted in an Arrhenius plot together
with the relaxation times $\tau_s(q,T)$ obtained from the incoherent scattering
functions $F_s(q,t)$ for different wave-vectors $q$ and the inverse of the 
diffusion constant $D(T)$~\cite{delgado_05a}.
This figure well elucidates the role of bonds and nodes in the 
fundamental difference between the high $T$ and 
the low $T$ regime of the relaxation dynamics: At low temperatures (i.e. 
$T \leq 0.1$) the bond lifetime sets the longest relaxation time scale in the 
system, and it becomes much longer than the simulation time window at the 
lowest $T$. That is, whereas at high temperatures the relaxation in the system 
is dominated by particle collision and diffusion, at the temperatures where the 
aggregation starts to produce persistent structures the main mechanism for the 
relaxation is the bond breaking.\\
From the figure, we recognize that at high $T$ the relaxation 
time $\tau_s(q,T)$ follows a $T^{-0.5}-$dependence for all $q$, which can 
easily be understood from the ballistic motion of the free particles and 
the small clusters. (Recall that in this $T-$range the distribution of the 
cluster size is independent of $T$~\cite{delgado_05a}.) 
For wave-vectors that are large, $q \geq 2.0$, the $T-$dependence of 
$\tau_s(q,T)$  at low temperatures is relatively weak but still somewhat 
stronger than $T^{-0.5}$. This can be understood by realizing that on the 
corresponding length scales the particles can still undergo an (almost) 
ballistic motion, despite the fact that they are, at low $T$, connected to 
other particles, since the whole local structure is moving ballistically. 
This type of motion is no longer possible if one considers length scales that 
become comparable to the size of the mesh of the network which is around 10 
and thus corresponds to a $q$ smaller than 1.0. 
For $q=0.55$ we find at low temperatures a $T-$dependence of $\tau_s$ that is 
close to an Arrhenius law whereas for smaller wave-vectors we find an even 
stronger $T-$dependence. On length scales that are thus comparable or larger 
than the typical mesh size of the network the $T-$dependence is thus very 
similar to the one characterizing the slow dynamics of dense glasses.\\
Also included in Fig.~\ref{fig3} is the inverse of the diffusion
constant $D$, scaled by a factor $A=49.2$ in order to make
it coincide with the value of $\tau_s(q=0.14,T=1.0)$. We find that in
the whole $T-$range investigated this quantity follows very closely
the $T-$dependence of $\tau_s(q,T)$ for small $q$, which is evidence
that in this system the relaxation of the structure is closely linked
to the diffusive motion of the particles, in contrast to the behavior
found in dense glass-forming systems in which dynamical heterogeneities
make that the $T-$dependence of $D$ is weaker than the one of
$\tau_s$~\cite{richert02}. This behavior corresponds to the lowest 
temperature dynamical regime associated to the presence of the 
persistent network.\\ 
Since here we would like to point out in particular  
the features of bond and nodes correlation at low $T$, we note that 
for $T \leq 0.1$, $\tau_b$ shows an Arrhenius dependence on the temperature, 
in agreement with the picture mentioned above. As shown in the figure, 
the lifetime of the bonds is around a factor of 20 larger than the one 
of the anchor points, but the $T-dependence$ is the same. 
This result is reasonable since the anchor particles experience on average 
more mechanical stress than the bridging particles and, due to their higher 
coordination, have less possibilities to yield to this stress.
Hence it is more likely that their bonds are broken. 
Fig.\ref{fig3} also shows that the low $T$ regime associated to the 
presence of the network, whose complex dynamics have been studied in 
Ref.\cite{delgado_05b}, corresponds to $\tau_b$ being the longest time scale 
and $\tau_{3b}$ becoming of the order of the longest relaxation times in the 
system. At these temperatures, there are practically no free particles 
(see Figs.\ref{fig1} and \ref{fig2}) and more than $97\%$ of them belong 
to the spanning network. Therefore one can envision the relaxation process 
in this system that the connecting branches of the network detach from 
the anchor particles, the branch reorients and attaches itself to a 
different branch, thus creating a new anchor point. 
Of course, the possibility that a branch breaks at a bridging particle 
can not be neglected completely since there are significantly more bridging 
particles than anchor particles. Thus this type of motion will contribute to 
the relaxation dynamics as well.
\begin{figure*}
\vspace*{1cm}
\begin{center}
\includegraphics[width=0.9\linewidth]{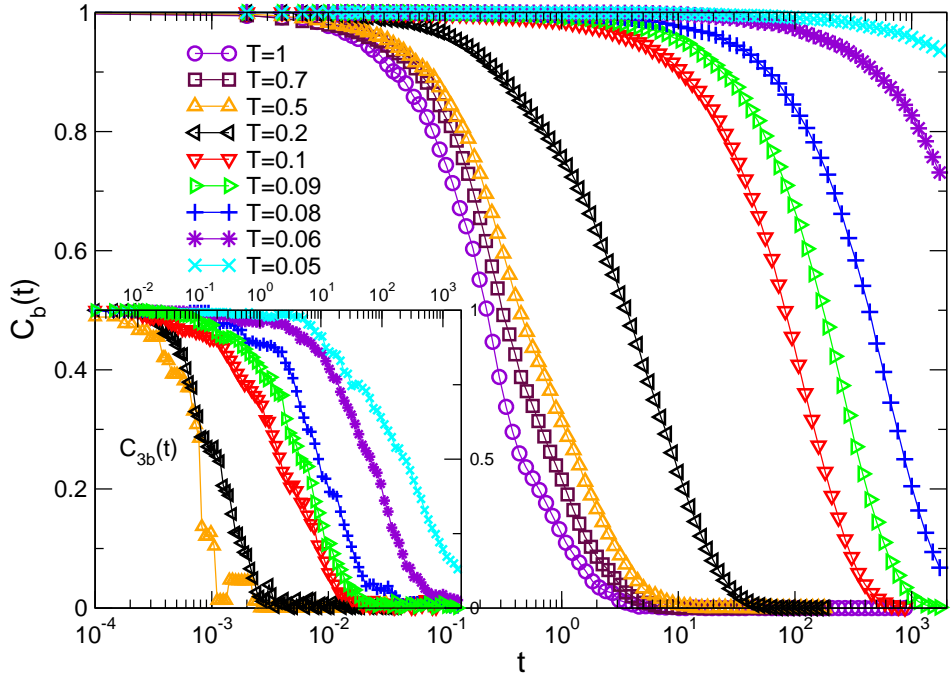}
\caption{
(Color online) {\bf Main panel}: Bond correlation function $C_b(t)$ for
different temperatures $T$. {\bf Inset}:
Time correlation function of the anchor points of the network, $C_{3b}(t)$,
for different temperatures $T$.
}
\label{fig4}
\end{center}
\end{figure*}
\\Let us now see how the time dependence of bond and nodes correlations at 
different temperature can be connected to the scenario just presented.
In the main panel of Fig.~\ref{fig4}, $C_b(t)$ is plotted as a function of
the time for different temperatures $T$. 
Although the long time decay of these functions is well described by a 
single exponential law at all the temperatures, we can actually distinguish 
two different regimes, one of high temperatures ($T > 0.1$) and the low 
temperature regime, corresponding to $T \leq 0.1$. 
At high temperature the particle collisions promote uncorrelated bond breaking 
or formation leading to a short time decay of bond correlation whose amount 
monotonically decreases with temperature. At long times the bond breaking 
due to the energy activation process will eventually lead to the complete 
decay of correlations.
At low temperatures instead, the energy activation process appears to be 
the only relevant process in the decay of bond correlation 
(due to the less and less important role of particle collisions).\\
In the inset of Fig.~\ref{fig4}, $C_{3b}(t)$ is plotted as a function of the 
time for different temperatures. At the higher temperatures ($T > 0.1$) at 
this volume fraction particles of connectivity $3$ are extremely rare, as 
already shown by Fig.~\ref{fig1}. As long as a spanning network is not 
formed, i.e. at temperatures $T\geq 0.06$, the long time decay of time 
correlation functions $C_{3b}(t)$ follows a simple exponential law, whose 
characteristic relaxation time increases with decreasing $T$. 
In addition to this regime, at temperatures when a persistent spanning 
network is present in the system, i.e. $T \le 0.055$, 
the decay of $C_{3b}(t)$ becomes also stretched, with a
stretching exponent $\beta$ decreasing with $T$ ($\beta \simeq 0.55$ at
$T=0.05$). This indicates that, once the network is formed and it is
sufficiently persistent, the breaking and formation of the nodes will be
associated not only to the overcoming of an activation energy but also to
some heterogeneous and cooperative dynamic process 
\cite{delgado_05b,delgado_05c}.\\
The scenario emerging from these data can be interestingly connected to the 
detailed analysis of the wave vector dependence of $\tau_s(q,T)$ presented in 
Ref.\cite{delgado_05b}. There we have discussed the striking 
differences between the high temperature and low temperature regime of the 
time correlations of particle displacement. 
At high temperatures, by decreasing the wave vector, 
$\tau_s(q,T)$ smoothly crosses over from a ballistic to a diffusive type of 
behavior with an exponential decay of time correlations.
At low temperatures, instead, once that a permanent spanning network is 
present, two clearly different regime, respectively at high and low wave 
vectors, can be individuated. We have been able to show that the relaxation 
at high wave vectors is due to the fast cooperative motion of pieces of the 
gel structure, whereas at low wave vectors the overall rearrangements of the 
heterogeneous gel make the system relax via a stretched exponential decay
of the time correlators. The coexistence of such diverse relaxation 
mechanisms is apparently determined by the formation of the gel network 
(i.e. the onset of the elastic response of the system) and it is 
characterized by a typical crossover length which is of the order of the 
network mesh size.\\

\section{Conclusion}
\label{conclu}
We have analyzed the structural and dynamical features in the gel formation
by molecular dynamics simulations of a new model for colloidal gels, based on
directional effective interactions. We have focused on the change in the 
topology corresponding to the formation of the gel network and on the time 
correlation of bonds and nodes of the gel network. 
Our analysis shows that, once that bonds are practically permanent over the 
simulation time window, the onset of the persistent network corresponds
to the onset of some heterogeneous and complex dynamics, which 
is directly related to the anchor points of the network.
It is evident that the anchor points are very important for the
mechanical stability of the structure. On length scales of the order of
the interparticle distance the structure is quite flexible whereas on
the scale of the mesh size of the network, which is given by the typical
distance between the anchor points, the structure becomes somewhat more
rigid due to the constrains imposed by the enhanced connectivity. However,
even on that length scale the system is relatively soft as compared to the
one found in dense glasses. With decreasing temperature the connecting
chains will become stiffer and hence give rise to an increased effective
interaction between the anchor points \cite{delgado_05b,delgado_05c}.

Acknowledgments:
Part of this work has been supported by the Marie Curie Fellowship
MCFI-2002-00573, the 
ANR-06-BLAN-0097-01 and EU Network Number
MRTN-CT-2003-504712.

\end{document}